\magnification=\magstep1
\input epsf.sty
\overfullrule=0pt

\def\today{\ifcase\month\or January\or February\or March\or
April\or May\or June\or July\or August\or September\or
October\or November\or December\fi \space\number\day,
\number\year}

\headline={\hfill{BOS - \today}}

\def\vp{{\varphi}}
\def\vep{{\varepsilon}}

\def\part{{\rm part}}

\def\IR{{\Bbb R}}

\def\frac#1#2{{#1\over#2}}

\def\IR#1{I\!\!R^{#1}}

\baselineskip=12pt

\centerline{{\bf{\underbar{Collapse of an Instanton}}}}
\vglue .5truein

\settabs\+ \phantom{.25truein} &Jagellonian University\qquad 
&L.D. Landau Institute\qquad 
&University of Toronto\hfill \cr

\+&P. Bizo\'n &Yu N. Ovchinnikov &I.M. Sigal\cr
\+&Jagellonian University &L.D. Landau Institute &University of Toronto\cr 
\+&Krak\'ow& Moscow& Toronto\cr

\vglue .5truein

\noindent {\bf \underbar{Abstract}}

{\bigskip\narrower We construct a two parameter family of 
collapsing solutions to the 4+1
Yang-Mills equations and derive the dynamical law of the collapse. 
Our arguments indicate that this family of solutions is stable.  
The latter fact is also supported by numerical simulations.\bigskip}

\noindent {\bf 1. \underbar{Introduction}}
\bigskip

Blow up problems for nonlinear Schr\"odinger, wave and heat
equations have been a subject of active research in the last 15
years (see [1,2,3] for reviews and [4,5] for recent papers on the
subject).  A further surge of interest in blow-up for nonlinear
wave equations has been recently motivated by their role in
attempting to understand the problem of singularity formation in
General Relativity (see [6] for a recent review). In this paper we
describe the asymptotic dynamics  of blowup for radial solutions
of the semilinear wave equation
$$
 \eqalignno{\ddot u &= \Delta u + \frac{1}{r^2} f(u), &(1.1)\cr}
 $$
in $\IR{2}$, where $u=u(t,r)$, 
$r$ is the radial variable and
$$
  \eqalignno{f(u) &=2 u (1- u^2). &(1.2)\cr}
$$
Our analysis is applicable to a wider class of ``double-well''
type of nonlinearities $f(u)$ producing kink-type solutions,
though some of such nonlinearities, e.g. wave maps nonlinearity
$f(u)=-{1\over 2}\sin (2u)$, lead to certain subtleties and will be 
considered elsewhere.
\medskip

Before stating the results we show how the equation (1.1) arises
and put the problem in a broader context. We consider Yang-Mills
(YM) fields in $d+1$ dimensional Minkowski spacetime
 (in the following Latin and Greek
indices take the values $1,2,\dots,d$ and $0,1,2,\dots,d$
respectively).
 The gauge potential $A_{\alpha}$ is
a one-form with values in the Lie algebra $g$ of a compact Lie
group $G$.
 In terms of the curvature
$F_{\alpha\beta}=\partial_{\alpha}A_{\beta}-\partial_{\beta}A_{\alpha}+
[A_{\alpha},A_{\beta}]$ the YM equations take the form
$$
\eqalignno{\partial_{\alpha}F^{\alpha\beta}+  [A_{\alpha},
F^{\alpha\beta}] & = 0, &(1.3)\cr}
$$
where $[,]$ is the Lie bracket on $G$. For simplicity, we take
here $G=SO(d)$ so the elements of $g=so(d)$ can be considered as
skew-symmetric $d\times d$ matrices and the Lie bracket is the
usual commutator. Assuming
 the spherically symmetric ansatz [7]
$$
\eqalignno{A^{ij}_{\mu}(x) &=
\left(\delta^i_{\mu}x^j-\delta^j_{\mu}x^i\right)
\frac{1-u(t,r)}{r^2}, &(1.4)\cr}
$$
 equations (1.3) reduce to the scalar semilinear wave equation for the
magnetic potential $u(t,r)$
$$
\eqalignno{\ddot u & = \Delta_{(d-2)} u + \frac{d-2}{r^2} u
(1-u^2), &(1.5)\cr}
$$
where $\Delta_{(d-2)}=\partial^2_{r} +\frac{d-3}{r}\partial_r$ is
the radial Laplacian in $d-2$ dimensions.
\medskip
 The central question for
equation (1.5) is:  can solutions starting from  smooth initial
data
$$
\eqalignno{ u(0,r) &= f(r), \qquad \dot u(0,r) = g(r) &(1.6)\cr}
$$
become singular in future?  An answer to this question depends
critically on the dimension $d$. To see why we recall two basic
facts. The first fact is the conservation of (positive definite)
energy
$$
\eqalignno{ E &= \int\limits_0^{\infty} \left({\dot u}^2 +{u'}^2
+\frac{d-2}{2 r^2} (1-u^2)^2\right) r^{d-3} dr. &(1.7)\cr}
$$
 The second fact
 is scale invariance of  the YM equations: if
$A_{\alpha}(x)$ is a solution of (1.3), so is $\tilde
A_{\alpha}(x)=\lambda^{-1} A_{\alpha}(x/\lambda)$, or
equivalently,  if $u(t,r)$ is a solution
  of (1.5), so is $\tilde
u(t,r)=u(t/\lambda,r/\lambda)$. Under this scaling the energy
scales as
 $\tilde E=\lambda^{d-4}
E$, hence the YM equations are subcritical for $d\leq 3$, critical
for $d=4$, and supercritical for $d\geq 5$. In the subcritical
case, shrinking of solutions to arbitrarily small scales costs
infinite amount of  energy, so it is forbidden by energy
conservation. This is a heuristic explanation of global regularity
of the YM equations in the physical dimension which was proved in
[8] and  [9]. In contrast,
 in
the supercritical case shrinking of solutions might be
energetically favorable and consequently singularities are
anticipated. In fact, for $d\geq 5$ equation (1.5) admits
self-similar solutions which are explicit examples of
singularities [10,11] and numerical simulations indicate that the
stable self-similar solution determines the universal asymptotics
of blowup for large initial data [12].
\medskip
 In the critical dimension
$d=4$ the problem of singularity formation
 is more subtle because the scaling argument is inconclusive.
In this case there are no smooth self-similar solutions, however
there is a family of static solutions $\chi(r/\lambda)$, where
$\lambda>0$ and
$$
 \eqalignno{ \chi(r)&=\frac{1-r^2}{1+r^2}. &(1.8)\cr}
$$
Using physicists' terminology we shall refer to this solution as
the instanton. Numerical simulations indicate that the existence
of the scale-free instanton plays a key role in the dynamics of
blowup, namely the blowup has the universal profile of the
instanton which shrinks adiabatically to zero size [12]. More
precisely, it was conjectured in [12] that near the blowup time
$t_*$ the solution has the form
$$
  \eqalignno{  u(t,r) &\approx
  \chi\left(\frac{r}{\lambda(t)}\right),&(1.9)\cr}
$$
where the scaling parameter $\lambda(t)$ tends to zero as
$t\rightarrow t_*$. A natural question is: what determines the
evolution of the scaling parameter; in particular, what is the
asymptotic behaviour of $\lambda(t)$ for $t\rightarrow t_*$? In
this paper we address this question and show that  for some
initial conditions that are close to the instanton
$$
\eqalignno{\lambda(t) &\sim \sqrt{\frac{2}{3}}\frac{t_*-t}
{\sqrt{-\ln(t_*-t)}} &(1.10)\cr}
$$
as $t\rightarrow t*$. The logarithmic correction to the
self-similar behaviour is characteristic for the blow-up in
critical equations - it implies that the speed of blow-up goes
asymptotically to zero and consequently no kinetic energy
concentrates at the singularity (for a different approach see
[13]).
\bigskip

\noindent {\bf 2. \underbar{Results}}
\bigskip

Thus we consider the intial value problem (1.1), (1.2) and
(1.6).  Since we consider radial solutions only the full
Laplacian $\Delta$ can be replaced by the radial Laplacian
$\Delta_r={1\over r}\partial_r r\partial_r$.
\medskip

Our main result states that if initial conditions (1.6) are
sufficiently close to
\break$\bigg(\chi\left(\frac{r}{\lambda_0}\right), -\frac{\dot
\lambda_0}{\lambda_0}
\frac{r}{\lambda_0}\chi'\left(\frac{r}{\lambda_0}\right)\bigg)$,
where $\lambda_0>0$  and $\dot \lambda_0<0$, then the resulting
solution is of the form
$$
\eqalignno{u(r,t)&=\chi\left(\frac{r}{\lambda}\right)+O(\dot\lambda^2),
&(2.1)\cr}
$$
\noindent where the scaling parameter
$\lambda=\lambda(t)$ satisfies  the following equation
$$
\eqalignno{\lambda\ddot \lambda&=\frac{3}{4}\dot
\lambda^4,&(2.2)\cr}
$$
\noindent with the initial conditions $\lambda_0$ and $\dot \lambda_0$
. In fact, our
procedure allows us to find the solution $u(x,t)$ to any
order in $\dot\lambda^2$ with the term of order $\dot\lambda^2$
given explicitly.
\medskip

Note that solutions of Eqn (2.2) with the initial conditions such
that  $\dot\lambda_0<0$, decrease to zero as $t\to t_*$ for some
$t_*$ with $|\dot\lambda|$ decreasing so that our approximation
improves as $t\to t_*$. This and Eqn (2.1) imply that the
instanton collapses as $t\to t_*$.
\medskip

To demonstrate the property of Eqn (2.2) mentioned above
 we note that Eqn (2.2)
can be integrated explicitly.  Indeed, rewrite (2.2) as
$\dot\lambda^{-3}\ddot\lambda=\frac{3}{4}\lambda^{-1}\dot\lambda$
and integrate the resulting equation to obtain
$$
\eqalignno{-\frac{1}{2}\dot\lambda^{-2}&=\frac{3}{4}
(\ln\lambda+\ln c),&(2.3)\cr}
$$
\noindent where $c>0$.  The latter equation can be rewritten as
$$
\eqalignno{\dot\lambda^2\ln\left(\frac{1}{c\lambda}\right)
&=\frac{2}{3}.&(2.4)\cr}
$$
\noindent This relation shows that we must have
$$
\eqalignno{c\lambda&<1.\cr}
$$
Using Eqn (2.4) we obtain the equation for $c$ in terms of
$\lambda_0$
$$
\eqalignno{\ln\left(\frac{1}{c\lambda_0}\right)
&=\frac{2}{3}{\dot\lambda_0}^{-2}.&(2.5)\cr}
$$

 \noindent We have two cases.
\medskip

\item{a.}$\dot\lambda_0>0$.  Then $c\lambda \uparrow 1$ and $\dot\lambda
\uparrow\infty \hbox{ as } t \hbox{ approaches some}\,\; t_*>0$.
Moreover, $c~\lambda~=~1-(\frac{3}{2}c^2)^{1/3}(t_*-t)^{2/3}$ as
$t\to t_*$.
\medskip
\item{b.}$\dot\lambda_0<0$.  Then $\dot\lambda<0$ for $t>0$ and
there is $t_*>0 \hbox{ s.t. }\lambda\to 0 \hbox{ as }t\uparrow
t_*$.  The value $t_*$ can be found from (2.4):
$$
\eqalignno{t_*&=\sqrt{\frac{3}{2}}\int\limits_0^{\lambda_0}
d\lambda\ln^{1/2} \left(\frac{1}{c\lambda}\right).&(2.6)\cr}
$$

\noindent Taking into account (2.5) this  gives
$$
\eqalignno{t_*& \approx \lambda_0 |\dot\lambda_0|^{-1}.&(2.7)\cr}
$$
The time $t_*$ is the point of collapse.
\medskip

Note that in this case the function $|\dot\lambda|$
decreases as $t\to t_* \hbox{ as
}\left[\ln\left(\frac{1}{c\lambda}\right) \right]^{-1/2}$ and
therefore our approximation improves as $t\to t_*$.
\medskip

Solutions of Eqn (2.2) with $\dot\lambda_0<0 \hbox{ and }
\lambda_0>0$ have the following asymptotics as $t\to t_*$
$$
\eqalignno{\lambda&=\sqrt{\frac{2}{3}}\frac{t_*-t}
{\sqrt{-\ln(t_*-t)}}.&(2.8)\cr}
$$

In conclusion we observe that Eqn (2.2) is invariant under the
transformation
$$
\lambda (t) \to \mu^{-1} \lambda (\mu t)
$$
inherited from the invariance of parent Eqn (1.1) under the
scaling transformation
$$
u(r,t) \to u (\mu r, \mu t).
$$
\bigskip
\vfil\eject

\noindent{{\bf 3. \underbar{Scaling transform and zero mode}}}
\bigskip

A key role in our derivation is played by the fact that Eqn (1.1)
is scale covariant under the transformation
$$
\eqalignno{u(r,t)&\to u(r/{\lambda},t/{\lambda}),&(3.1)\cr}
$$
\noindent i.e. if $u(r,t)$ is a solution to (1.1), then so is
$u(r/{\lambda},t/{\lambda})$. In particular, if $v(r)$ is a
stationary solution, then so is $v(r/{\lambda}),\lambda>0$.  The
infinitesimal change of the instanton $\chi$ under this
transformation is
$$
\eqalignno{\chi&\to\chi+\delta\lambda\zeta,&(3.2)\cr}
$$
\noindent where the function $\zeta$ is defined by
$$
\eqalignno{\zeta (r)&:=\partial_\lambda|_{\lambda=1}\chi
(r/{\lambda})=-r\partial_r\chi(r).&(3.3)\cr}
$$
\noindent Explicitly
$$
\eqalignno{\zeta(r)&=\frac{4r^2}{(1+r^2)^2}\,.&(3.4)\cr}
$$
\medskip

Of course, $\zeta$ is the zero mode,
$$
\eqalignno{L\zeta&=0,&(3.5)\cr}
$$
of the linearization of the r.h.s. of (1.1) on $\chi$, i.e. of the
operator (recall, $\Delta_r={1\over r}\partial_r r\partial_r$)
$$
\eqalignno{L&:=-\Delta_r-\frac{1}{r^2}f'
\big(\chi(r)\big).&(3.6)\cr}
$$
\noindent (This operator is the
variational or Fr\'echet derivative, $L=\partial\phi(\chi)$,
 of the map
$ \phi(u)=-\Delta u-\frac{1}{r^2}f(u) $ at the instanton $\chi$.)
\medskip

\noindent The following properties of the operator $L$
will be important for us:
\medskip
\item{-}$L=L^*\ge 0$
\medskip
\item{-}$L$ has a simple eigenvalue at 0 with the
eigenfunction $\zeta$
\medskip
\item{-}the continuous spectrum of $L$ fills $[0,\infty)$.
\medskip
The first and third properties are obvious 
and the second property follows
from the equation $L\zeta=0$ and the fact that $\zeta>0$ by the
Perron-Frobenious Theory (see [14,15]). 
\medskip

Consider solutions of Eqn
(1.1) of the form $u(r,t):=v(r/\lambda, t)$, where $\lambda>0$
depends on $t$.  Plugging the function $u(r,t)=v(r/\lambda,t)$
into Eqn (1.1), we obtain the following equation for $v$ and
$\lambda$:
$$
\eqalignno{\Delta_yv+y^{-2}f(v)&=-\dot\lambda^2 B_1
v-\lambda\ddot\lambda B_2 v +\lambda^2\partial_t^2 v-2
\lambda\dot\lambda B_2\partial_t v,&(3.7)\cr}
$$
\noindent where
$$
\eqalignno{B_1&=-y\partial_y-(y\partial_y)^2\qquad \hbox{and}
\qquad B_2=y\partial_y.&(3.8)\cr}
$$
\bigskip

\noindent{{\bf 4. \underbar{Orthogonal decomposition}}}
\bigskip

We look for a solution of Eqn (1.1) of the form
$$
\eqalignno{u(r,t)\equiv v(r/{\lambda}, t)&=\chi(r/{\lambda})+
w(r/{\lambda},t)&(4.1)\cr}
$$
\noindent with $\lambda=\lambda(t)$ and
$$
\eqalignno{w(y,t)\qquad &\hbox{ small for all times.}&(4.2)\cr}
$$
\noindent Moreover, to fix the splitting between the dynamics
of $\lambda$ and of $w$ we require that $w$ is orthogonal
to the zero mode $\zeta$:
$$
\eqalignno{\int\limits_0^{\infty}\zeta(y)
w(y,t)ydy&=0.&(4.3)\cr}
$$
\noindent The last two conditions will
give us the dynamic law for $\lambda$.
\medskip

Now we plug the decomposition $v(y,t)=\chi(y)
+w(y,t)$ into Eqn (3.7) and use that the function $\chi$ satisfies
the equation
$$
\eqalignno{\Delta_y\chi+y^{-2}f(\chi)&=0,&(4.4)\cr}
$$
\noindent to obtain the equation for $w$:
$$
\eqalignno{(L+\lambda^2\partial^2_t)w&=F(w,\lambda),&(4.5)\cr}
$$
\noindent where, recall, $L=L_\chi$ is the linearized operator
around $\chi $
$$
\eqalignno{L&:=-\Delta_y-y^{-2}f'(\chi)&(4.6)\cr}
$$
\noindent and
$$
\eqalignno{F(w,\lambda)&:=\dot\lambda^2B_1(\chi+w)
+\ddot\lambda\lambda B_2(\chi+w)\cr
&\qquad+y^{-2}N(w)+2\lambda\dot\lambda
B_2\partial_t w &(4.7)\cr}
$$
\noindent with the
nonlinearity $N(w)$ defined by
$$
\eqalignno{N(w)&:=f(\chi+w)-f(\chi)-f'(\chi)w,&(4.8)\cr}
$$
\noindent which in the case $f(u)=2(1-u^2)u$ gives
$$
\eqalignno{N(w)&=-6\chi w^2-2w^3.&(4.9)\cr}
$$
\bigskip
\noindent{{\bf 5. \underbar{Perturbative analysis. Outline}}}
\bigskip

We explain the main idea of our approach by proceeding 
formally and ignoring infrared divergences arising in
an attempt to justify our analysis.  In the next section we
present a full perturbation theory.  We look for
a solution to Eqn~(4.5) in the form
$$
\eqalignno{w(y,t)&=\sum\limits_{j\geq 1} \dot\lambda^{2j}\xi_j(y).&(5.1)}
$$
\noindent Plugging this expansion into Eqn(4.5) we
arrive at a series of equations 
$$
\eqalignno{L\xi_j&=F_j(\xi_0,\dots,\xi_{j-1}),&(5.2)\cr}
$$
$j\geq 1$, where $\xi_0=\chi$.
\medskip

We demonstrate our approach by analyzing the cases $j=1$ and 2
in detail.  We begin with $j=1$.  It is clear from (4.5)--(4.7) that
$$
\eqalignno{F_1(\xi_0)&=B_1\chi.& (5.3)\cr}
$$
\noindent Thus $\xi_1$ satisfies the equation
$$
\eqalignno{L\xi_1&=B_1\chi.&(5.4)\cr}
$$
\noindent Since, as we show in Appendix 1,
$$
\eqalignno{\int\zeta B_1\chi&=0,&(5.5)\cr}
$$
\noindent Eqn (5.4) has a solution.  The general solution
of this equation is
$$
\eqalignno{\xi_1(y)&=\xi_{10}(y)+\alpha_1\zeta(y)&(5.6)\cr}
$$
\noindent for any $\alpha_1\in\IR{}$.  Here
$$
\eqalignno{\xi_{10}(y)&=-{y^4\over (1+y^2)^2}&(5.7)\cr}
$$
\medskip

Now plugging $w=\dot\lambda^2\xi_1+O(\dot\lambda^4)$ into
(4.3), 
and using (5.5) and $B_2\chi=-\zeta$, we obtain
$$
\eqalignno{\int\zeta\left(\dot\lambda^4B_1\xi_1
-\ddot\lambda\lambda\zeta
-\dot\lambda^46y^{-2}\chi\xi_1^2\right)&=O(\dot\lambda^6)+O(\lambda
\ddot\lambda\dot\lambda^2).&(5.8)\cr}
$$
\noindent It will be shown in Appendix 1 that the coefficients
in front of $\alpha_1$ and $\alpha_1^2$ (remember (5.6)) vanish:
$$
\eqalignno{\int\zeta B_1\zeta-12\int\zeta y^{-2}\chi\xi_{10}\zeta&=0
&(5.9)\cr}
$$
and
$$
\eqalignno{\int \zeta y^{-2}\chi\zeta^2&=0&(5.10)\cr}
$$
\noindent Therefore relation (5.8) becomes
$$
\eqalignno{\ddot\lambda\lambda-\gamma\dot\lambda^4&=
O(\dot\lambda^6),&(5.11)\cr}
$$
\noindent where
$$
\eqalignno{\gamma&=\int\zeta\left(B_1\xi_{10}
-6y^{-2}\chi\xi_{10}^2\right)\left(\int\zeta^2\right)^{-1}.&(5.12)\cr}
$$
\noindent We show in Appendix 1 that $\gamma=3/4$ which, in
the leading order, brings us to Eqn(2.8):
$$
\eqalignno{\lambda\ddot\lambda&={3\over 4}\dot\lambda^4.&(5.13)\cr}
$$
\noindent As (5.11) shows, this equation is valid modulo the
correction $O(\dot\lambda^6)$.
\medskip

Now we proceed to the second term in (5.1) and derive a
correction to Eqn(5.11) (or(5.13)).  Remember that $\xi_2$
is defined by (5.2) with $j=2$.  Keeping in mind Eqn(5.13)
we have
$$
\eqalignno{F_2(\xi_0,\xi_1)&=B_1\xi_1
+\gamma B_2\chi-6y^{-2}\chi\xi_1^2.&(5.14)\cr}
$$
\noindent By (5.8) we have that
$$
\eqalignno{\int\zeta F_2(\xi_0,\xi_1)&=0&(5.15)\cr}
$$
\noindent so that the equation $L\xi_2=F_2(\xi_0,\xi_1)$ is
solvable. Now plugging
$$
\eqalignno{w&=\dot\lambda^2\xi_1+\dot\lambda^4\xi_2
+O(\dot\lambda^6)&(5.16)\cr}
$$
into (4.3), we obtain the equation for $\lambda$
$$
\eqalignno{\lambda\ddot\lambda-\gamma\dot\lambda^4
+\delta\dot\lambda^6+\vep\dot\lambda^2\lambda\ddot\lambda
+O(\dot\lambda^8)&=0&(5.17)\cr}
$$
with $\delta$ and $\vep$ given in terms of integrals of
$\xi_1$ and $\xi_2$ which can be explicitely computed.
Observe that Eqn(5.17) is equivalent to the equation
$$
\eqalignno{\lambda\ddot\lambda-\gamma\dot\lambda^4
+(\delta-\gamma\vep)\dot\lambda^6+O(\dot\lambda^8)&=0.&(5.18)\cr}
$$
\noindent We can continue in the same manner to find the
equation for $\lambda$ to an arbitrary order in $\dot\lambda^2$.
\medskip

Though the perturbation theory outlined above leads (as
we will see in the next section) to correct---in the 
leading order---equations for the dilation parameter
$\lambda$, it is, in fact, inconsistent.  The leading
correction, $\xi_1$, does not vanish at infinity and 
consequently the resulting solution has infinite energy.
Worse, higher-order corrections grow at infinity.  Moreover,
orthogonality condition (4.3) is not applicable (and
as a result the parameter $\alpha_1$ in (5.6) cannot be 
determined).  The reason for this inconsistency is that
the term $-\lambda^2\partial^2_tw$ cannot be treated as
a perturbation at large distances.  A correct perturbation
theory taking into account the leading contribution of
this term at infinity is presented in the next section.
\bigskip

\noindent{{\bf 6. \underbar{Perturbative analysis}}}
\bigskip

In this section we justify formal analysis of Section~5.
We look for a solution, $w$, of Eqn (4.5) in the form
$$
\eqalignno{w(y,t)&=\sum\limits_{j\geq 1}\dot\lambda^{2j}\xi_j(y)
\vp_j(\dot\lambda^4y^2,t).&(6.1)\cr}
$$
\noindent We fix the functions $\xi_j$ and $\vp_j$ by requiring
that (a) $\xi_j$ and $\vp_j$ are of the order $O(1)$,
(b) the functions $\vp_j$ satisfy the relations 
$$
\eqalignno{\vp_j(z,t)&=1\quad {\rm for}\quad z\ll 1&(6.2)\cr}
$$
and
$$
\eqalignno{\vp_j(z,t)&=\vp_j(z)+O(\dot\lambda^2),&(6.3)\cr}
$$
\noindent (c) the following equations are satisfied
$$
\eqalignno{(L+\lambda^2\partial^2_t)
(\dot\lambda^{2j}\xi_j\vp_j)&=\dot\lambda^{2j}F_j,&(6.4)\cr}
$$
where the functions $F_j$ are $O(1)$ and depend only 
on $\xi_0\vp_0,\dots,\xi_{j-1}
\vp_{j-1}$ with $\xi_0=\chi$ and $\vp_0\equiv1$:
$$
\eqalignno{F_j&\equiv F_j(\xi_0\vp_0,\dots,\xi_{j-1}
\vp_{j-1})&(6.5)\cr}
$$
\noindent and (d) the functions $\xi_j(y)$ satisfy the equations
$$
\eqalignno{L\xi_j&=F_j(\xi_0,\dots,\xi_{j-1}).&(6.6)\cr}
$$
\noindent As will be shown below these requirements
will define the functions $\xi_j$ and
$\vp_j$ uniquely, at least in the leading order.
\medskip

We demonstrate our approach by analyzing the cases $j=1$ and 2
in detail.  We begin with $j=1$.  It is clear from (4.5)--(4.7) that
$$
\eqalignno{F_1(\xi_0\vp_0)&=B_1\chi.& (6.7)\cr}
$$
\noindent Thus $\xi_1$ satisfies the equation (5.4).  Recall
that due to (5.5) the latter equation is solvable and its general
solution is given by (5.6).
The constant $\alpha_1$ in (5.6) is determined from the condition
$$
\eqalignno{\int\zeta\cdot\xi_1\vp_1&=0.&(6.8)\cr}
$$
Since it plays no role in what follows we do not compute it here
(see, however, (6.23) below).
\medskip

Now plugging $w=\dot\lambda^2\xi_1\vp_1+O(\dot\lambda^4)$ into
(4.3), omitting $\vp_1$ (justification for this will be provided
later) and using (5.5) and $B_2\chi=-\zeta$, we obtain (5.8) which 
as shown in Section~5 leads to (5.11) with $\gamma=3/4$.
\medskip

Now we return to expansion (6.1) and find the equation for
$\vp_1$. Recall that $\vp_1$ is defined through equations
(6.2)--(6.6) with $j=1$. We derive from these equations the
equation for $\vp_1(z,t)$ in the leading order in $\dot\lambda^2$
and in the domain $y\gg 1$. To this end we use Eqn (5.13) to
estimate the order of higher derivatives of $\lambda$. In the
leading order we can ignore the dependence of $\vp_1(z,t)$ on $t$.
Using Eqns (6.4) with $j=1$ and Eqn (6.7) and using that for $y\gg
1$
$$
\eqalignno{\xi_1&=-1+O\left({1\over y^2}\right),\quad
y\partial_y\xi_1=-{4\over y^2}+O\left({1\over y^4}\right),\quad
B_1\chi=-{4\over y^2}+O\left({1\over y^4}\right),&(6.9)\cr}
$$
\noindent we obtain after lengthy but elementary computations
that $\vp_1$ satisfies the equation
$$
\eqalignno{z^2\partial^2_z\vp_1+\left(z+\gamma z^2\right)\partial_z\vp_1
-\left(1-{1\over 2}\gamma z\right)\vp_1&=-1.&(6.10)\cr}
$$
\noindent To this we add the boundary conditions
$$
\eqalignno{\vp_1(0)&=1.&(6.11)\cr}
$$
\noindent The second boundary condition, $\vp_1'(0)$, or, alternatively,
an arbitrary constant in the general solution to (6.10)--(6.11), is found
by matching the solution to (6.10)--(6.11) in the region $z\ll 1$
with solution to (6.4) with $j=1$ and (6.7) in the region $y\gg1$.
This is done in Appendix 2 where it is also shown that
$$
\eqalignno{\vp_1(z)&=\cases{
    1-{\gamma z\over4}\left(\ln{z\over\dot\lambda^4}-\frac{7}{3}\right)
	&for $z\ll1$\cr
    &\cr
    \frac{c}{\sqrt z}+\frac{2}{\gamma z}+
	{\bar c\over\sqrt z}e^{-\gamma z}&for $z\gg1$\cr}
&(6.12)\cr}
$$
\noindent for some constants $c$ and $\bar c$. Thus we have
$$
\eqalignno{\dot\lambda^2\xi_1(y)\vp_1(\dot\lambda^4y^2)&=O(y^{-1})
\quad {\rm for}\quad\dot\lambda^2 y\gg1.&(6.13)\cr}
$$
\noindent This implies, in particular, that the integral in
(6.8) converges and it gives
$$
\eqalignno{\alpha_1&=O\left(\ln{1\over\dot\lambda^2}\right).&(6.14)\cr}
$$
\medskip

Now we proceed to the second term in (6.1) and derive a
correction to Eqn(5.11) (or(5.13)).  Remember that $\xi_2$
is defined by (6.6) with $j=2$.  Keeping in mind Eqn(5.13)
we choose
$$
\eqalignno{F_2(\xi_0\vp_0,\xi_1\vp_1)&=B_1(\xi_1\vp_1)
+\gamma B_2\chi-6y^{-2}\chi(\xi_1\vp_1)^2.&(6.15)\cr}
$$
\noindent By (5.8) we have that
$$
\eqalignno{\int\zeta F_2(\xi_0,\xi_1)&=0&(6.16)\cr}
$$
\noindent so that the equation $L\xi_2=F_2(\xi_0,\xi_1)$ is
solvable.  Eqns(6.2)--(6.6) with $j=2$ imply an equation for
$\vp_2$ which is analyzed in a similar way as the equation for
$\vp_1$.  Now plugging
$$
\eqalignno{w&=\dot\lambda^2\xi_1\vp_1+\dot\lambda^4\xi_2\vp_2
+O(\dot\lambda^6)&(6.17)\cr}
$$
into (4.3) and setting $\vp_1$ and $\vp_2$ to 1, we
obtain the equation (5.17) (or (5.18)) for $\lambda$.
We can continue in the same manner to find the
equation for $\lambda$ to an arbitrary order in $\dot\lambda^2$.
\bigskip

\centerline{\underbar{{\bf Conclusion}}}
\bigskip

We found, perturbatively, a two parameter family of
collapsing solutions (parametrized as $\lambda_0$
and $\dot\lambda_0$) to the
nonlinear wave equation (1.1)--(1.2) arising from 
the Yang-Mills equation in 4+1 dimensions.  We also
found the corresponding dynamics of collapse.  The
perturbation theory developed suggests
that this family is (asympototically)
stable. This conclusion is supported
by numerical simulations we performed (some 
of the results of these simulations are given in
the figure below).
\bigskip

\centerline{\underbar{{\bf Acknowledgements}}}
\bigskip

Research on this paper was supported 
by KBN under grant 2P03B00623,  
by a RFFI grant and
by NSERC under grant N7901. 
Two of the authors (P.B. and I.M.S)
are grateful to the ESI, Vienna, for hospitality in July 2001
when the work on this paper started.
\bigskip

\centerline{\underbar{{\bf Appendix 1 Computation of
various integrals}}}
\bigskip

In this appendix we show (5.9),
(5.14a), (5.14b) and $\gamma=3/4$ (see~Eqn~(5.17)).
\medskip

\item{1.} (5.9). Recall $\zeta=
-y\partial_y\chi(y) \hbox{ and }
B_1=-y\partial_y-(y\partial_y)^2$. Hence $B_1\chi
=(1+y\partial_y)\zeta$ and
$$
\int\zeta B_1\chi=
\int\zeta(\zeta+y\zeta')=\int\limits_0^{\infty}
y\zeta(y\zeta)'dy.
$$
Integrating by parts we get
$$
\int\limits_0^{\infty}
y\zeta(y\zeta)'dy=\frac{1}{2}(y\zeta)^2\bigg|^{\infty}_0,
$$
and since the boundary term vanishes, we get
$\int\zeta B_1\chi=0$. Nota bene, this shows that the
orthogonality condition for $j=1$ is basically equivalent to the
square integrability of $\zeta$.
\medskip
In what follows we use the following relation ($m\leq n-2$)
$$
\int\limits_0^{\infty} \frac{x^m dx}{(1+x)^n} = \frac{m
(m-1)\cdot...\cdot 1}{(n-1)(n-2)\cdot...\cdot(n-m-1)} .
$$
\item{2.}(5.14a).  Show that
$$
\int\zeta B_1\zeta-12\int\zeta y^{-2} \chi w_1\zeta = 0 .
$$
\noindent Compute
$$
\eqalign{\int\zeta B_1\zeta&=-\int\limits_0^{\infty}
(\zeta\zeta'y^2+\zeta \partial_y (y\zeta')y^2 dy\cr
&=\int\limits_0^{\infty}(\zeta\zeta'y^2+\zeta'^2y^3)dy\cr
&=\int\limits_0^{\infty}(-\zeta^2y+\zeta'^2y^3)dy.\cr}
$$ \noindent
This gives
$$
\eqalignno{{1\over 8}\int\zeta B_1\zeta
	&=2\int\limits_0^{\infty}\left(\frac{-y^4}{(1+y^2)^4}
		+\frac{4y^4(1-y^2)^2}{(1+y^2)^6}\right)ydy\cr
	&=\int\limits_0^{\infty}\left(\frac{-x^2}{(1+x)^4}
		+\frac{4x^2(1-x)^2}{(1+x)^6}\right)dx\cr
	&=-\frac{2\cdot1}{3\cdot2\cdot1}
		+\frac{4\cdot2\cdot1}{5\cdot4\cdot3}
		-\frac{8\cdot3\cdot2\cdot1}{5\cdot4\cdot3\cdot2}+
		\frac{4\cdot4\cdot3\cdot2\cdot1}
		{5\cdot4\cdot3\cdot2\cdot1}={1\over5 },}
$$
$$
\eqalign{{1\over8}\int\zeta^2y^{-2}\chi w_1
	&=-2\int\limits_0^{\infty}\frac{y^4\cdot y^{-2}}{(1+y^2)^4}
		\frac{1-y^2}{1+y^2}\frac{y^4}{(1+y^2)^2}ydy\cr
	&=-\int\limits_0^{\infty}\frac{x^3(1-x)}{(1+x)^7}dx\cr
	&=-\frac{3\cdot2\cdot1}{6\cdot5\cdot4\cdot3}
		+\frac{4\cdot3\cdot2\cdot1}
		{6\cdot5\cdot4\cdot3\cdot2}={1\over 60}.} 
$$
\noindent Hence $\displaystyle{\int\zeta
B_1\zeta-12\int\zeta y^{-2} \chi w_1\zeta = 8
(\frac{1}{5}-\frac{12}{60})=0}$.

\item{3.}(5.14b). Compute
$$
\eqalignno{{2\over {4^3}}\int\zeta^3y^{-2}\chi
	&=2\int\limits_0^\infty\frac{y^4}{(1+y^2)^6}
		\frac{1-y^2}{1+y^2}ydy\cr
	&=\int\limits_0^{\infty}\frac{x^2(1-x)}{(1+x)^7}dx\cr
	&=\frac{2\cdot1}{6\cdot5\cdot4}-\frac{3\cdot2\cdot1}
		{6\cdot5\cdot4\cdot3}=0.\cr}
$$
\medskip

\item{4.}$\gamma=3/4$. Compute
$$
\eqalign{{1\over8}\int\zeta^2&=2\int\limits_0^{\infty}
\frac{y^4}{(1+y^2)^4}ydy\cr
&=\int\limits_0^{\infty}\frac{x^2dx}{(1+x)^4}\cr
&=\frac{2\cdot1}{3\cdot2\cdot1}={1\over3 }.\cr}
$$
\noindent
Compute
$$
\int\zeta B_1w_1-6\int\zeta y^{-2}\chi w_1^2=2,
$$
\noindent so $\gamma={3\over4}$.
\medskip

\centerline{\bf\underbar{Appendix 2 Solution $\vp_1$}}
\medskip

In this appendix we find the solution to `initial' value problem
(5.19)--(5.20) matching the solution to (5.4) with $j=1$ in
the region $1\ll y\ll\dot\lambda^{-2}$.  In the region
$\{z\ll1\}$ Eqns (5.19)--(5.20) have the general solution
$$
\eqalignno{\vp_1&=1-{\gamma\over 4}\ln z(z+\dots)+c'{\gamma\over 4}
z(1+\dots)&(A2.1)\cr}
$$
\noindent with an arbitrary constant $c'$.
\medskip

For $z\gg 1$ Eqn (5.19) has the general solution
$$
\eqalignno{\vp_1&={2\over\gamma z}+c{1\over\sqrt z}+\bar c
{e^{-\gamma z}\over\sqrt z}&(A2.2)\cr}
$$
\noindent with arbitrary constants $c$ and $\bar c$.  Here
${1\over\sqrt z}$ and ${e^{-\gamma z}\over \sqrt z}$ are
solutions of the corresponding homogeneous equation in the
region $z\gg1$.
\medskip

It remains to find the constant $c'$ in (A2.1).  To this end we
match $\vp_1(\dot\lambda^4y^2)$ (in the leading order) to the
solution of the equation
$$
\eqalignno{\left(L+\lambda^2{\partial^2\over\partial t^2}\right)w
&=\dot\lambda^2 B_1\chi&(A2.3)\cr}
$$
\noindent in the region $1\ll y\ll \dot\lambda^{-2}$.  We
find the solution of the latter in the leading order in $\dot\lambda^2$
by a pertubation theory:
$$
\eqalignno{w&=\dot\lambda^2 w_1+\dot\lambda^6w_2+\dots,&(A2.4)\cr}
$$
\noindent where $w_1=\xi_1$ (see Eqn(5.6)).
This implies the equation for $w_2$:
$$
\eqalignno{Lw_2&=2\gamma\xi_{10}.&(A2.5)\cr}
$$
\noindent Two solutions of the corresponding homogeneous
equation are (see (3.5))
$$
\eqalignno{\eta_1&={y^2\over(1+y^2)^2}\quad {\rm and}\quad
\eta_2={y^2\over 4}+{3\over2}-{13\over 4(y^2+1)}-{1\over
4y^2(y^2+1)}+{3y^2\ln y^2\over(y^2+1)^2}.&(A2.6)\cr}
$$
\noindent By the method of variation of constants we obtain
$$
\eqalignno{w_2&=c_1\eta_1+c_2\eta_2&(A2.7)\cr}
$$
\noindent where the functions $c_1$ and $c_2$ are given by
$$
\eqalignno{c_1&=-\gamma\Bigg\{{y^4\over 8}+y^2-
{4\over y^2+1}+{1\over(y^2+1)^2}-{y^6\ln y^2\over (y^2+1)^3}&\cr
&\quad-{3y^4\ln y^2\over 2(y^2+1)^2}-{3y^2\ln y^2\over y^2+1}
+3\int\limits_0^{y^2}{ds\ln s\over s+1}+4\Bigg\}&(A2.8)\cr}
$$
and
$$
\eqalignno{c_2&=\gamma\left\{\ln(y^2+1)+{3\over y^2+1}
-{3\over 2(y^2+1)}+{1\over 3(y^2+1)^3}-{11\over 6}\right\}.&(A2.9)\cr}
$$
\noindent Eqns (A2.7)--(A2.9) for $y\gg1$ yield
$$
\eqalignno{w_2&=-{\gamma\over 8}y^2
+\gamma{y^2\over 4}\ln(1+y^2)+\dots&\cr
&={\gamma\over 4}y^2\ln y^2+\dots&(A2.10)\cr}
$$
\noindent Since on the other hand
$
w_1=-1+O(y^2)
$,
\noindent we find in $y\gg 1$ that
$$
\eqalignno{w&=-\dot\lambda^2\left[1-{\gamma\over 4}z\ln z+\dots\right]
&(A2.11)\cr}
$$
\noindent where $z=\dot\lambda^4y^2$.  Comparing (A2.11) with (A2.1)
we find
$$
\eqalignno{c'&=\ln y^2-\ln(\dot\lambda^4y^2)
=\ln\dot\lambda^{-4}&(A2.12)\cr}
$$
\noindent and therefore
$$
\eqalignno{\vp_1(z)&=1-{\gamma\over 4}z\ln{z\over\dot\lambda^4}+\dots
\quad {\rm for}\quad z\ll 1.&(A2.13)\cr}
$$
\vfil\eject
\bigskip
\centerline{\epsfbox{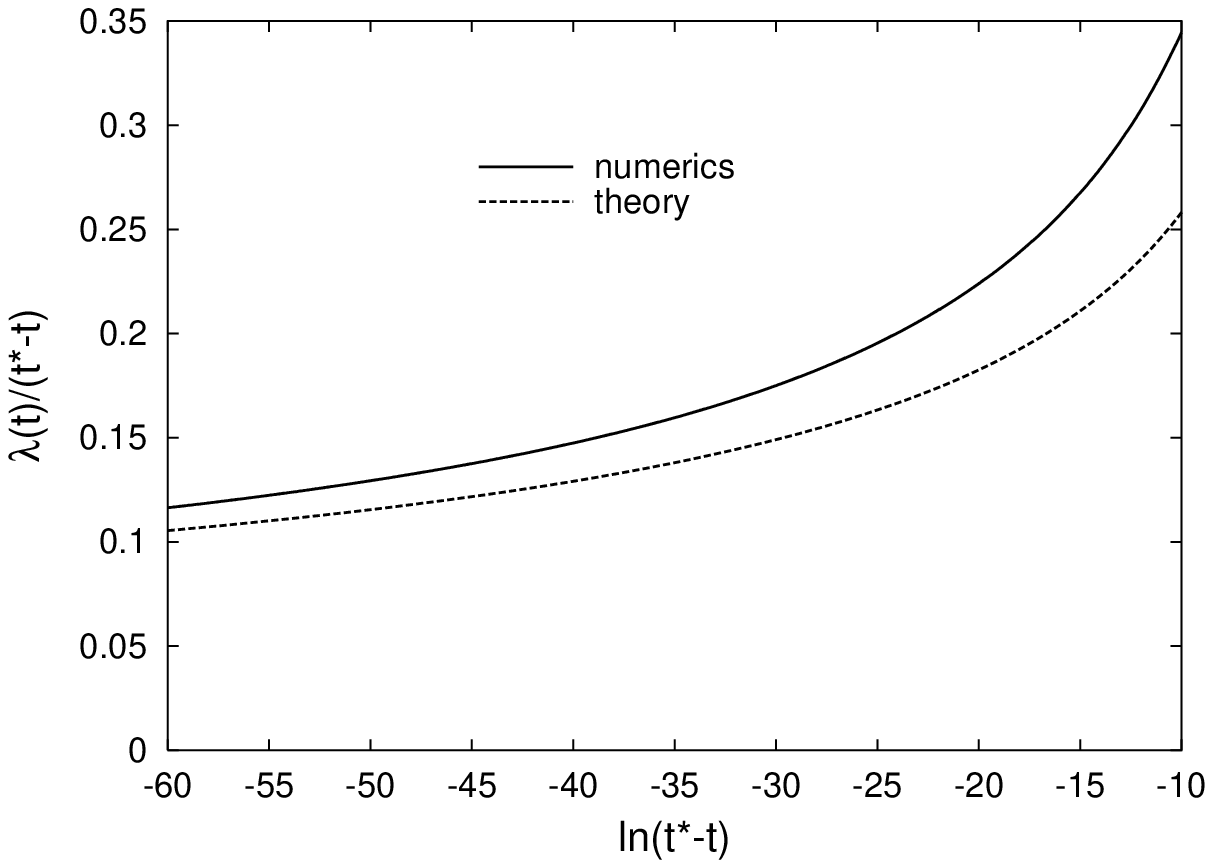}}
\bigskip
\noindent Comparison of the numerically computed scaling parameter
divided by $t^*-t$ with the analytic formula
$\frac{\lambda(t)}{t^*-t}=\sqrt{\frac{2}{3}}
(-\ln(t^*-t))^{-1/2}$.
 Note that there are no free parameters to be fitted. We believe that the apparent discrepancy
 (which is of the order of
 10\%  at $\ln(t^*-t)=-60$)  can be accounted for by including  higher order corrections to the formula
 (1.10).

 \centerline{} \vfil\eject \centerline{REFERENCES}
\baselineskip=12pt

\medskip
\bigskip

\item{[1]} Singularities in Fluids, Plasmas and Optics, 
vo. 404 of Nato Advances Study Institute, Series C: 
Mathematical and Physical
Sciences, R.E. Calfish and G.C. Papanicolaou, eds, Kluwer Academic
Publishers, 1993
\medskip

\item{[2]} L. Berg\'e, 
Wave collapse in physics: principles and
applications of to light and plasma waves, 
{\it Physics Reports} 303
(1998) 259-370.
\medskip

\item{[3]} C. Sulem and P.-L. Sulem,
Nonlinear Schr\"odinger equation: self-focusing and wave collapse,
{\it Springer}, 2000.
\medskip

\item{[4]} M.P. Brenner, P. Constantin, L.P. Kadanoff, A. Schenkel
and S.C.Venkataramani, 
Diffusion, attraction and collapse,
{\it Nonlinearity} 12 (1999) 1071-1098.
\medskip

\item{[5]} Yu. N. Ovchinnikov and I.M. Sigal, 
Multiparameter family of collapsing solutions 
for a critical nonlinear Schr\"odinger equation, 
{\it Preprint}, 2002.
\medskip

\item{[6]}
 Alan D. Rendall,  
Applications of the theory of evolution equations to 
general relativity,
in Proceedings of GR16, (Eds.) N. T. Bishop and S. D. Maharaj,
{\it World Scientific}, 2002.
\medskip

\item{[7]} O. Dumitrascu, {\it Stud. Cerc. Mat.} 34(4), 329 (1982).
\medskip

\item{[8]} D. Eardley and V. Moncrief, {\it Commun. Math. Phys.} 83, 171
(1982).
\medskip

\item{[9]} S. Klainerman and M. Machedon, {\it Ann. Math.} 142, 39
(1995).
\medskip

\item{[10]} T. Cazenave, J. Shatah, and A. Shadi Tahvildar-Zadeh,
{\it Ann. Inst. Henri Poincare} 68, 315  (1998).
\medskip

\item{[11]} P. Bizo\'n, {\it Acta Phys. Polon.} B33, 1893 (2002).
\medskip

\item{[12]} P. Bizo\'n and Z. Tabor, {\it Phys. Rev.} D64, 121701
(2001).
\medskip

\item{[13]} J. M. Linhart and L. A. Sadun, {\it Nonlinearity} 15, 219
(2002).
\medskip

\item{[14]} Yu.N. Ovchinnikov and I.M.~Sigal, 
Ginzburg-Landau equation I. Static Vortices, 
{\it CRM Proceeding and Lecture Notes}, Vol. 12, 199-219
(1997)
\medskip

\item{[15]} I.M.~Sigal,
Perron Frobenius Theory and Symmetry of solutions to 
nonlinear PDE's,
{\it Letters in Math. Phys.} 53, 313-320 (2000).

\bye